# Intermediate states of superconducting thermometers for x-ray microcalorimeters

H. Pressler, M. Koike, and M. Ohkubo[a)]
*Photonics Research Institute, National Institute of Advanced Industrial Science and Technology, 1-1-1, Umezono, Tsukuba, Ibaraki, 305-8568, Japan*

D. Fukuda, Y. Noguchi, M. Ohno, H. Takahashi, and M. Nakazawa
*Department of Quantum Engineering and System Sciences, University of Tokyo, 7-3-1, Hongo, Bunkyo-ku, Tokyo, 113-8656, Japan*

Intermediate states near normal-superconducting transition at ~100mK, important for a sensitive thermometer in x-ray calorimetry, have been studied by an imaging technique based on Low Temperature Scanning Synchrotron Microscopy (LTSSM). In conjunction with iridium superconducting transition-edge sensors operating in electrothermal feedback mode, the LTSSM visualizations reveal that a normal-superconducting phase separation takes place in the actual intermediate states.

[a)] Author to whom correspondence should be addressed; electronic mail: m.ohkubo@aist.go.jp

---

Microcalorimeters with superconducting thermometers, so-called transition-edge sensors (TESs), are actively studied because of superior high energy-resolution for energy dispersive spectroscopies.[1] Recent advance in the microcalorimeters with TESs operating in strong negative electrothermal feedback (ETF) mode gains the advantages of self-regulated temperature-bias and flexibility in device design to achieve a fast response time or a high photon-counting rate. The TES devices measure a photoabsorption-induced temperature rise by a resistance change due to sharp normal-superconducting transitions. The intermediate states in the middle of transitions have been studied experimentally or theoretically for many years,[2] and the results were included in the framework of TES operation.[3] Apart from an ideal intermediate case, actual superconducting transitions are influenced by such factors as magnetic field, current, phase-slip center, geometry, thermal conductivity, etc. Since the sensitivity of the TES devices influences the performance of microcalorimeters to a great extent, the investigation on the actual intermediate states of TESs in operation is very important.

One of the outcomes of the considerations on the intermediate states is the possibility of phase separation of normal and superconducting regions.[3] However, there are no reports on spatially-resolved measurements, and the phase separation has never been confirmed experimentally. We have employed a Low Temperature Scanning Synchrotron Microscope (LTSSM),[4] which enables 2D imaging with an x-ray microbeam.

Iridium TES devices with no additional absorbers were fabricated by rf-sputtering deposition and anisotropic etching of a silicon substrate covered by a $SiN_x$ membrane layer.[5] The iridium films on the $SiN_x$ membrane with an area of 1.4x1.4mm$^2$ and a thickness of 300nm have a size of 500x500μm$^2$ and a thickness of 50nm. The thermal conductance ($G$) between the TES device and the heat bath and the heat capacity ($C$) of the iridium film were estimated to be 0.75nW/K and 0.6pJ/K, respectively. The properties of the particular iridium film reported in this letter are as follows; resistivity at 300K $\rho_{RT}$=16μΩcm, residual resistivity $\rho_r$ =11μΩcm (residual resistance: $R_r$ =2.15Ω), superconducting transition temperature $T_c$ =134.5mK, transition width $\Delta T_c$= 1.3mK. The inset of Fig. 1 shows the top view of the TES device with the 250μm-wide and 150nm-thick niobium leads. The sample was mounted on a dilution refrigerator cold finger, achieving a base temperature of 55mK. A magnetic shield with an attenuation factor of 1000 and a niobium tube were used to eliminate the effect of vortex trapping. The iridium film was biased at constant voltages with a shunt resistance of 10mΩ. The TES device was scanned by a 3keV x-ray microbeam with a diameter of 20μm through x-ray windows. Reduction of the current flowing through the TES device was induced by individual photoabsorption events, and read out by a superconducting quantum interference device (SQUID) amplifier.[6] The signal pulses were recorded together with the coordinates of the x-ray microbeam. The spatial profiles of such current-pulse parameters as pulse height, risetime, and falltime were measured at different bias points between zero resistance and the normal resistance.

Figure 1 shows the spatial profiles measured at a bias voltage of 2.4μV. This bias point was in the vicinity of the zero resistance state. The current-pulse integral profile in Fig. 1 (a) is relatively flat within the area of the iridium film compared with those of other pulse parameters. The pulse integral is related to the absorbed photon energy ($E$) by a formula, $E=V\int\Delta I(t)dt$, when the pulse duration is much shorter than the intrinsic time constant ($C/G \approx$1ms) and the energy is removed by the reduction of Joule heating only. The slight nonuniformity is caused by a heat escape to the heat bath, since the falltime values in the left half region are not negligibly shorter than the intrinsic time constant. In contrast to Fig. 1 (a), the profiles of the pulse height, risetime, and falltime exhibit a large spatial nonuniformity along the $x$ direction parallel to the current flow. The pulse height profile has maximum values near the right end, while the risetime and falltime profiles are totally opposite and have the minimum values at the same position as the maximum pulse height. The best energy resolution of 14eV, which was limited by a baseline noise, was obtained near a maximum position. It is noteworthy that the spatial variation along the $y$ direction is small. This observation indicates that the phenomena can be treated in one dimension.

The mechanism responsible for the nonuniformity shown in Fig. 1 has been investigated by measuring bias-voltage dependence of the pulse height profiles along the $x$ direction. It is clearly visible in Fig. 2 that as the bias voltage increases, the positions of the pulse height maximums shift from right to left. The measured resistance values of the iridium film are plotted against the maximum pulse height positions in the inset of Fig. 2. The perfect linear relation except the data points near 500μm demonstrates that the normal region at the right of the maximum positions grows with increasing bias voltage, and the normal-superconducting boundaries are just at the maximum positions. In Fig. 1, the normal-superconducting boundary is indicated by the dashed lines, where the pulse height is maximum, while the risetime is minimum. A similar experiment with a Low Temperature Scanning Electron Microscope and a theoretical analysis were performed for superconducting bridges on bulk substrates.[7] For the present device, a theoretical analysis is in progress, and will be published elsewhere.

It is worth noting that the sides of the normal and superconducting regions are easily interchanged, as shown in Fig. 2. The solid and dashed lines for the bias at 3.3μV, which were recorded in different runs, are in perfect symmetry. The symmetry displays the interchange of the normal and superconducting sides. It has also been observed that a single photoabsorption event often initiates the interchange. The temperature rise for a 3keV photon absorption is estimated to be 0.8mK, when we assume that the heat immediately spreads all over the iridium film with a heat capacity of 0.6pJ/K. It is reasonable that the local temperature rise at the initial stage is considerably larger than the superconducting transition width of 1.3mK. The interchange of the normal and superconducting regions can be initiated by this hot spot in the superconducting region. This idea is supported by the fact that the absorption events in the superconducting regions produce additional spike signals prior to the ordinary current signals. The spike pulses cause the false short risetime in the pulse shape analysis, and appear as the holes near the left end in Fig. 1 (c). The spike pulses may be produced by a supercurrent-assisted hot spot formation mechanism, similarly to what occurs in superconducting single-photon optical detectors. [8]

The normal-superconducting phase separation is maintained by the Joule heating in the normal region and a temperature variation in the film. When we take the Wiedemann-Franz thermal conductance, $G=24.5T/R_n$[nW/K], and the Joule heating power, $P$, of 29pW in the ETF mode, the worst-case temperature difference across the film, $P/G$, is estimated to be 20mK that is significantly larger than the superconducting transition width.[3] Therefore, although the actual difference is surely smaller than 20mK, it is reasonable that the normal and superconducting regions coexist in the film. A photoabsorption event heats the film up locally and the heat spreads around the hot spot. When the heat reaches at the normal-superconducting boundaries, the current reduction signal is produced by a growth of the normal region. In this scenario, the maximum pulse height and minimum risetime are expected for photons absorbed just at the boundaries, where the temperature is equal to $T_c$. This is in good agreement with the observations.

The normal-superconducting phase separation is also confirmed by comparing the signal height produced in the normal region with that produced in the superconducting region. Since the curve for the bias at 5.4μV in Fig. 2 corresponds to the state where the normal-superconducting boundary is just at the middle of the film, it is an ideal case for the pulse height comparison. The heat capacity influences the current peak values by $\Delta T=E/C$. Therefore, the lower pulse height in the superconducting region is explained from the second order transition of superconductivity with a discontinuity of specific heat at $T_c$, which is expected to be as large as $\Delta c/c_n=1.43$, where $c_n$ is the normal-state electronic heat capacity per unit volume. The observed pulse height at the right end is 2.1 times as large as that at the left end. This ratio is consistent with the specific heat discontinuity.

The exact temperature variation profiles in the iridium film is unknown at this stage, but quasi-quantitative analysis is possible. As is shown in Fig. 2, the maximum pulse height decreases with increasing the bias voltage. The pulse height at the normal-superconducting boundaries in first approximation may be expressed by [7]

$$\Delta I = -\frac{V}{R_0^2}\Delta R = -\frac{V}{R_0^2}\frac{\rho_r}{S}\frac{1}{dT/dx}\Delta T \qquad (1)$$

where $\Delta R$ is the resistance increase due to the photoabsorption and proportional to the boundary shift $\Delta x$, $S$ is the cross section of the film, $\Delta T$ is the temperature rise due to the photoabsorption, $R_0$ is the resistance at the bias voltage $V$, and $dT/dx$ is the temperature gradient at the boundary. Since the position dependence of $\Delta T$ is expected to be negligible, $\Delta R$ is inversely proportional to $dT/dx$. The voltage drop due to a stray inductance and a pick-up coil inductance cannot be ignored, so that the bias voltage changes during the events. Nevertheless, we believe that the following preliminary analysis assuming the constant voltages provides a correct scheme. Figure 3 shows the bias-voltage dependence of the $\Delta R$ values that are obtainable from the applied bias voltages, the measured $R_0$, and the measured $\Delta I$ values. It is clear that the temperature gradient decreases with increasing bias voltage. The temperature profiles, which are deduced from the normal-superconducting boundary positions and the $dT/dx$ values assuming $\Delta T$ of 3mK, are shown in the inset of Fig. 3.

In conclusion, the LTSSM results have demonstrated that the normal-superconducting phase separation takes place in the TES device, as suggested in Ref. 3. The increase of the bias voltage results in the shift of the normal-superconducting boundary from the one end to the other. It has been elucidated that the interchange of the sides of normal and superconducting regions can easily occur in different measurement runs or be triggered by a single photoabsorption event. The temperature profiles across the film have been deduced from the normal-superconducting boundary positions and the temperature gradients. Since the phase separation influences the performance of the calorimeters, the LTSSM imaging technique can play an important role in investigating the detector physics and optimizing its performance.

We are grateful to M. Ukibe, T. Zama, M. Ataka, and the members of the electron acceleration facilities at AIST for the TES device fabrication, the LTSSM experiments, and fruitful discussion. Special thanks are due to F. Hirayama for the SQUID amplifier fabrication, and M. Koyanagi and N. Kobayashi for continuous encouragement from the inception of this study.

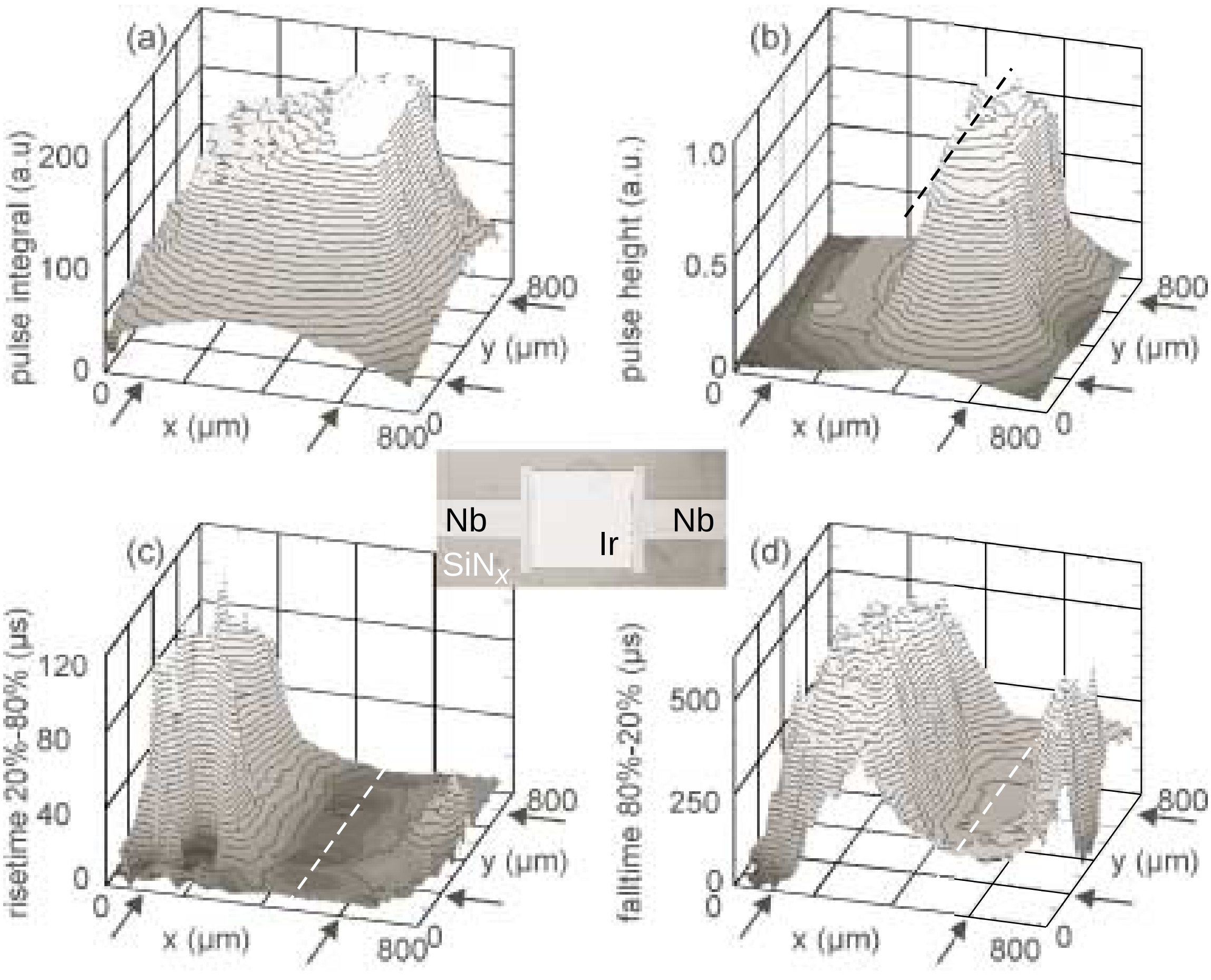

Fig.1 Spatial profiles of current signal parameters for 3keV x-rays at a bias voltage of 2.4μV; (a) pulse integral that is related to the photon energy, (b) pulse height, (c) risetime, and (d) falltime. The arrows show the perimeter of the iridium film. The dashed lines indicate the position of the normal-superconducting boundary. The inset shows the top view of the device.

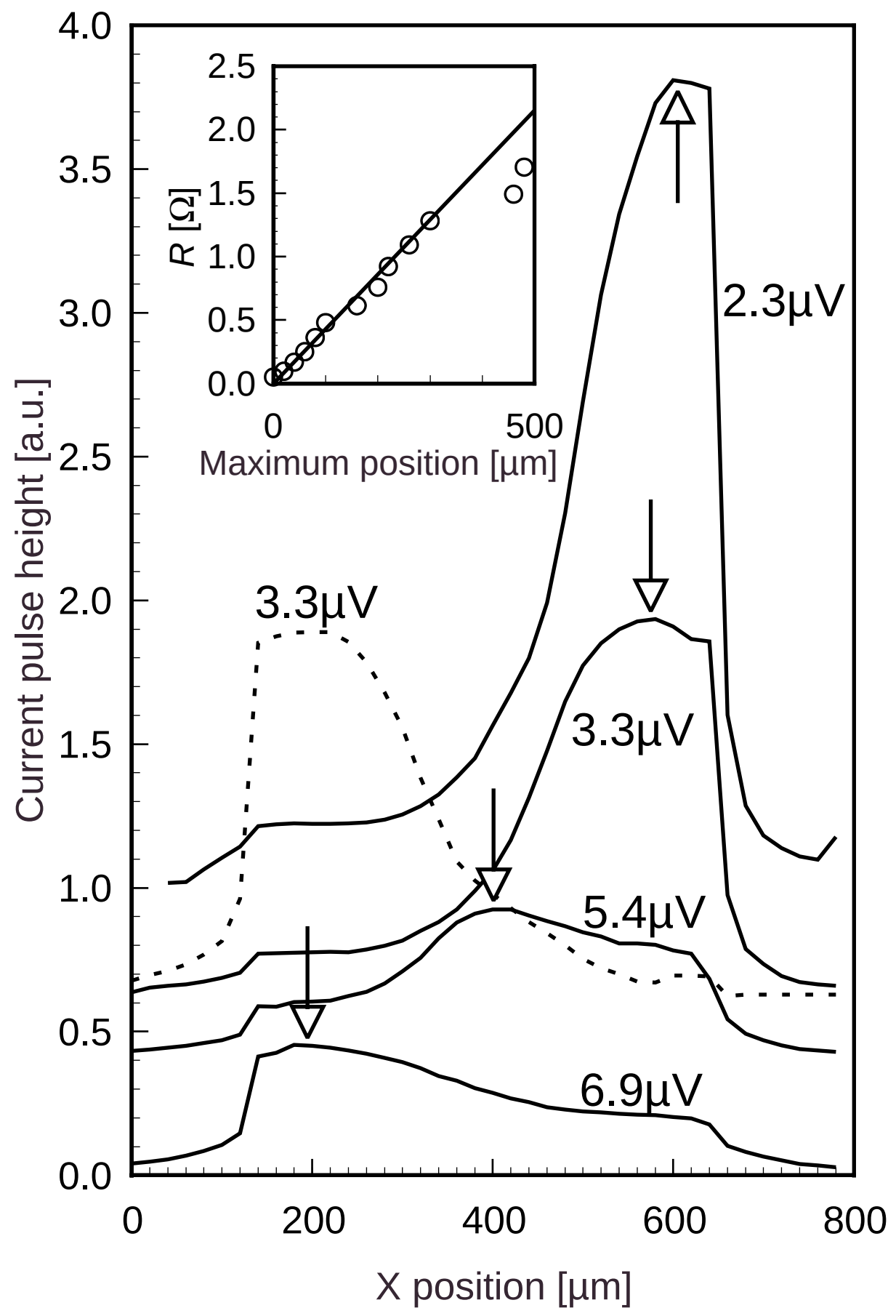

Fig. 2 Bias-voltage dependence of the current-pulse-height profiles along the *x* direction that is the current flow direction. For clarity, the curves are vertically shifted by 0.4 for 5.4μV, 0.6 for 3.3μV, and 1.0 for 2.3μV. The arrows indicate the normal-superconducting boundaries. The dashed curve for the bias at 3.3μV was measured in a different run. In the inset, the TES resistance values are plotted against the maximum positions of the pulse height profiles with the solid line that is expected from the growth of the normal region with a 11μΩcm.

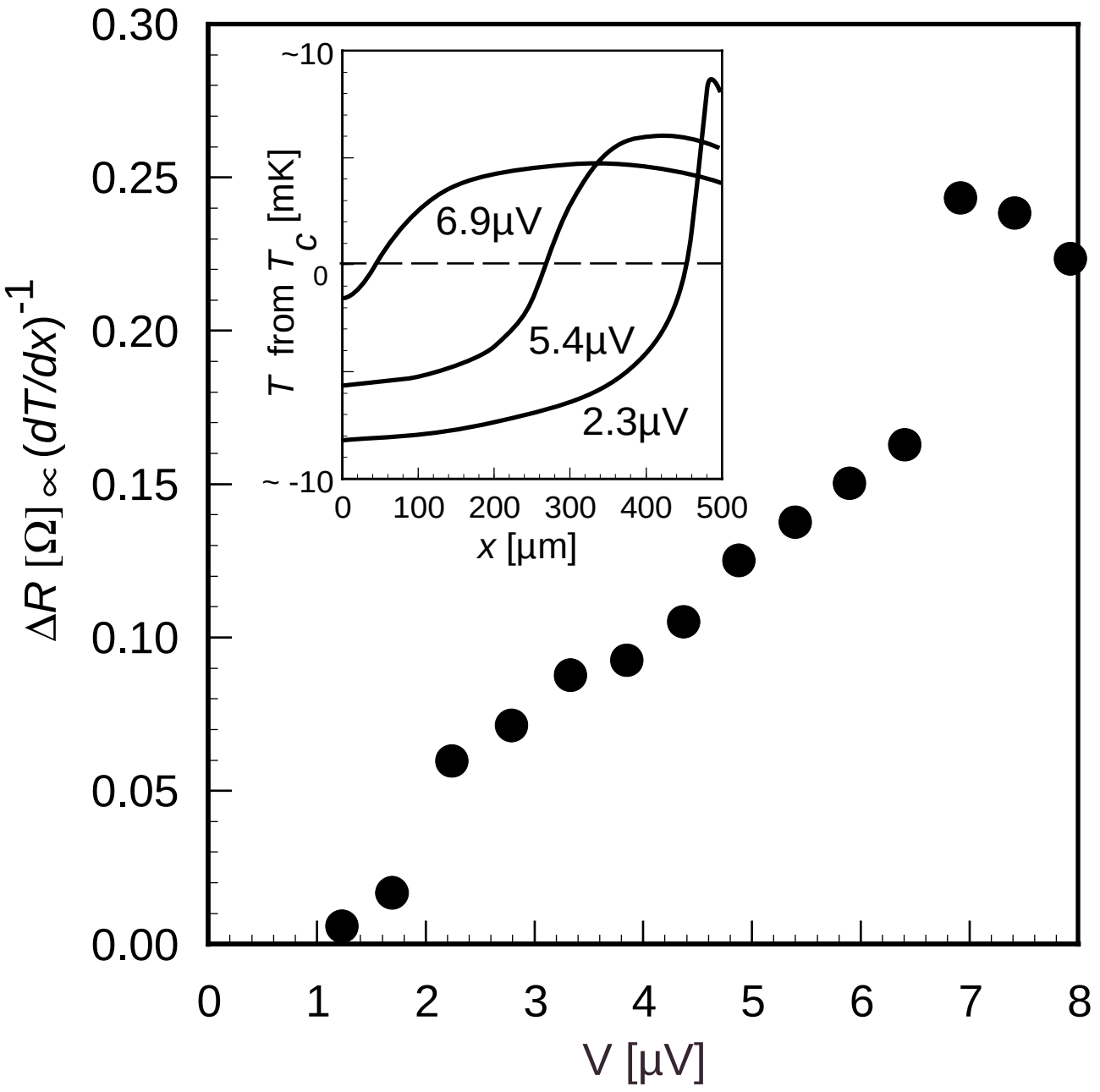

Fig. 3 Resistance change induced by a 3keV photon absorption event as a function of bias voltages. The inset is the temperature variation profiles deduced from the normal superconducting boundary positions and the temperature gradient values at the boundaries.